\documentclass[twocolumn,groupedaddress,amsmath,amssymb,aps,pra]{revtex4-2}

\usepackage{graphicx}
\usepackage{dcolumn}
\usepackage{bm}
\usepackage{hyperref}
\usepackage{physics}
\usepackage{siunitx}
\usepackage{color}

\newcommand{\iden}{\hat{\mathbb{I}}}
\newcommand{\superket}[1]{\left.\left|#1\right>\hspace{-2pt}\right>}

\begin{document}

\title{Effective qubit dephasing induced by spectator-qubit relaxation}

\author{P.~Jurcevic}
\thanks{Both authors contributed equally to this work. \\ petar.jurcevic@ibm.com \\ lcggovia@ibm.com}
\author{L.~C.~G.~Govia}
\thanks{Both authors contributed equally to this work. \\ petar.jurcevic@ibm.com \\ lcggovia@ibm.com}
\affiliation{IBM Quantum, IBM T.~J.~Watson Research Center, Yorktown Heights, NY 10598}

\date{\today}

\begin{abstract}
In many leading architectures for quantum computing, it remains to be understood if we can equate single-qubit coherence times measured in isolation, to that of coherence times measured in multi-qubit devices.
On a multi-qubit superconducting circuit platform we show an increase in the dephasing rate of a control qubit due to the spontaneous relaxation of spectator qubits coupled to the control qubit. We attribute this increased dephasing to random in time $Z$-phase kicks on the control qubit due to the interplay between spectator relaxation and the control-spectator $ZZ$-interaction. We measure the magnitude of this extra dephasing using Ramsey decay experiments, show how it can be corrected via dynamical decoupling pulse sequences, and demonstrate that randomized benchmarking is insensitive to the effect. Our experimental results are supported by a robust theoretical model that captures an arbitrary number of spectator qubits, and gives a simple, intuitive picture for the mechanism behind the enhanced dephasing.
\end{abstract}

\maketitle

\section{Introduction}

The physical architectures behind quantum computing platforms have shown tremendous improvement in recent years, such that qubit coherence times in leading platforms are not the only driver of performance for small systems. However, with growing system size and circuit complexity, it is important to understand if coherence times measured for single qubits in isolation faithfully represent the coherence times during multi-qubit circuits. This will be particularly important for developments towards quantum error correction \cite{Egan21,Chen21,Postler2021,Satzinger21,Ryan-Anderson21,Abobeih21,Chen21IBM,Krinner21,Zhao21} and quantum memories \cite{Bradley19,Anderson22}, where some qubits are likely to experience long idle times in an unknown quantum state.

In this work, on a superconducting qubit platform we experimentally demonstrate an enhancement in the effective dephasing rate of a control qubit due to the energy relaxation of excited spectator qubits that couple to the control qubit via a longitudinal ($ZZ$) interaction. The origin of this enhanced dephasing is the random phase-kicks on the control qubit associated with the random (in time) relaxation of the spectators, which in the ensemble average manifests as dephasing. We refer to this effect as spectator-decay-induced dephasing (SDID). As the error is coherent phase evolution on a shot-by-shot basis, it can be corrected using spin-echo sequences on the control qubit, which we also demonstrate.

The SDID mechanism is analogous to models for dephasing in superconducting qubits due to thermal two-level systems (TLSs) \cite{Matityahu16,Lisenfeld16}, and the decoherence induced by spin-spin interaction with spontaneous emission/absorption observed in trapped ion arrays \cite{Foss-Feig13,Bohnet16}. The distinction is that unlike these models, the state of the spectator qubits does not fluctuate, but exhibits at most one decay event for each qubit. Dephasing due to the SDID mechanism was also previously reported in nuclear magnetic resonance experiments \cite{Morton08,Gumann14}, and the impact of SDID was recently observed in superconducting qubits \cite{Livingston21,Tripathi21}. In the context of previous work, this manuscript represents a thorough experimental and theoretical exploration of the effect, including the derivation of an accurate theoretical model for an arbitrary number of spectator qubits. We note that during preparation of this manuscript we became aware of Ref.~\cite{McDonald21}, which presents general analytical solutions for the dissipative Ising model. This includes the SDID model we present in this work, and our theoretical solutions are in full agreement with those of Ref.~\cite{McDonald21}.

This manuscript is organized as follows. In section \ref{sec:SDID} we qualitatively explain the mechanism that leads to the SDID effect. In sections \ref{sec:Ramsey}, \ref{sec:cpmg}, and \ref{sec:RB} we present our experimental and theoretical results of studying SDID during Ramsey decay, composite pulse sequences, and randomized benchmarking. Finally, in section \ref{sec:Conc} we make our concluding remarks.

\section{Spectator-decay-induced dephasing}
\label{sec:SDID}

To understand the origin of the effect we refer to as spectator-decay-induced dephasing, we consider a single qubit (the control) coupled to one or more other qubits (the spectators) via a longitudinal or $ZZ$-interaction, as described by the Ising-model Hamiltonian
\begin{align}
    \hat{H} = \sum_{j=1}^N \nu_j \hat{Z}_0\otimes\hat{Z}_j.
\end{align}
Here $\nu_j$ is the $ZZ$-coupling strength between control qubit 0 and spectator qubit $j$, we consider $N$ spectator qubits, and we use $\hat{Z}_j$ as shorthand for the $(N+1)$-qubit operator with $\hat{Z}$ on qubit $j$ and identity on all others.

For spectator qubits in a static state, the control qubit simply evolves under the Hamiltonian $\hat{H}_0 = \nu_{s}\hat{Z}_0$, with $\nu_{s} = \sum_j^N (-1)^{s_j+1}\nu_j$ where $s_j = 0, 1$ denotes the state of the $j$'th spectator qubit. After some time $t$, this evolution imparts a $Z$-phase on the control qubit given by $\phi_t = \nu_st$. If at this point in time an initially excited spectator qubit relaxes to the ground state, $s_j = 1 \rightarrow 0$, then the magnitude of the $Z$-phase evolution of the control qubit will also change $\nu_s \rightarrow \nu'_s$. Thus, for a total evolution time $t'$, the total accumulated phase will be $\phi_\tau = \nu_st + \nu'_s(t'-t)$. The dynamics are completely coherent for any single run (or shot) of an experiment sensitive to this control-qubit phase evolution. However, since the time of spectator relaxation is a random variable, so too is the total phase accumulated. In the ensemble average of many such shots, this shot-by-shot random phase accumulation manifests as an effective dephasing of the control qubit state.

\section{Ramsey Decay}
\label{sec:Ramsey}

\begin{figure*}[ht!]
    \centering
    \includegraphics[width=0.99\textwidth]{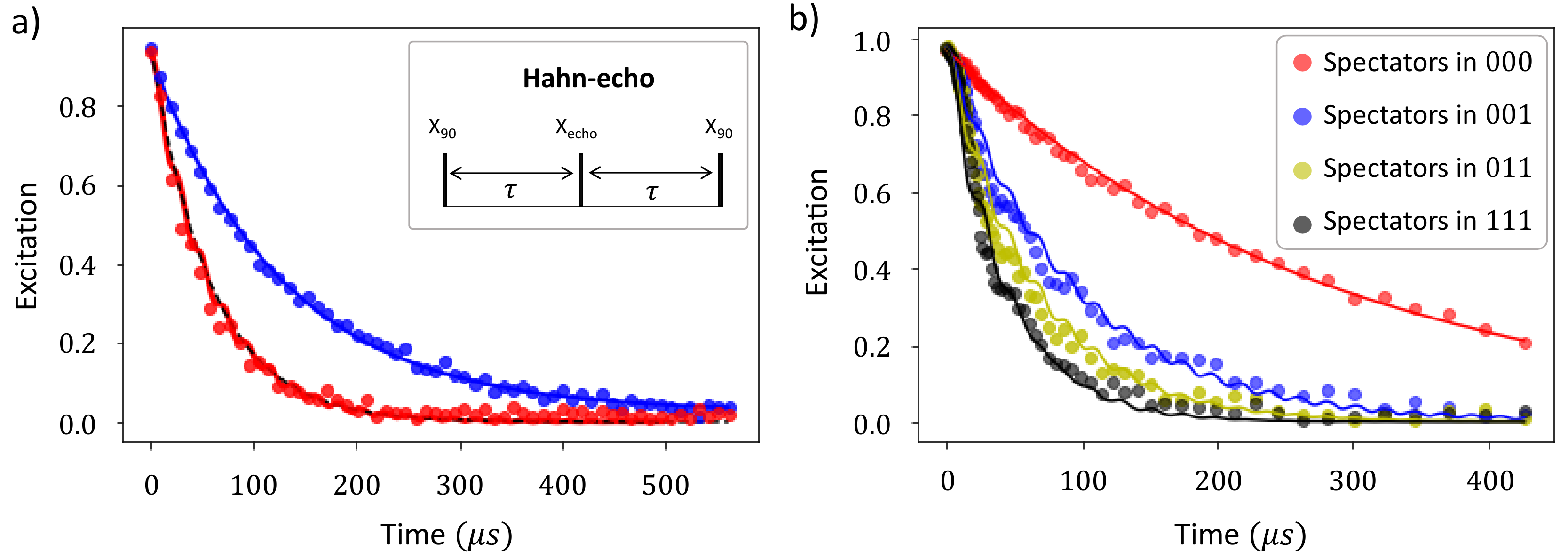}
    \caption{a) Ramsey spectroscopy on a control qubit with one $ZZ$-coupled spectator qubit. Inset: Ramsey echo sequence used to measure SDID. Blue dots: Experimental data with spectator prepared in $\ket{0}$. Red dots: Experimental data with spectator prepared in $\ket{1}$. Solid blue line: Exponential fit to dephasing data extracting $T^{(0)}_2=\SI{127}{\micro\second}$. Red solid line: Theory prediction based on Eq. (\ref{eqn:2q_tr}). Black dashed line: Exponential decay based on a heuristically adjusted decay rate given by $1/T_2^{(0)}+1/T_1^{(1)}$. b) SDID measured on a control qubit with three $ZZ$-coupled spectators. Dots are experimental data, red solid line is a exponential fit to extract the non-SDID dephasing rate and the remaining solid lines are theory curves. Red: All three spectators are prepared in the $\ket{0}$-state. Blue: One spectator is prepared in $\ket{1}$ while the others remain in $\ket{0}$. Yellow: Two spectators are prepared in $\ket{1}$ and Black: All three spectators are prepared in $\ket{1}$.}
    \label{fig1}
\end{figure*}

One of the dominant methods to measure phase accumulation is a Ramsey experiment \cite{Ramsey1950}. Here, we implement a Ramsey sequence with a Hahn-echo in the middle of the sequence on qubits from an IBM Quantum Falcon R5 device, see Fig.~\ref{fig1}a) inset. As we will explain later, a single Hahn-echo does not reduced sensitivity to the SDID effect, but it does entirely suppress unwanted frequency beating due to charge noise \cite{Riste13} and frequency miscalibrations, which would otherwise convolute the analysis. In Fig.~\ref{fig1} we compare the dephasing rate of a control qubit, $Q_0$, coupled by a $ZZ$-interaction of strength $4\nu\approx\SI{45}{\kilo\hertz}$ to a spectator qubit $Q_1$, with the spectator qubit prepared in either the $\ket{0}$-state (blue) or in the $\ket{1}$-state (red). The SDID-unaffected control qubit $T_2^{(0)}=\SI{127}{\micro\second}$ is measured (blue dots) and fit with a single exponential decay (solid blue line). The spectator qubit $T_1^{(1)}=\SI{107}{\micro\second}$ is determined with a separate experiment. After the spectator qubit is prepared in the excited state we see a significant reduction in the dephasing time of the control qubit (red dots). 

A theory model solely based on $ZZ$-coupling strength, control qubit dephasing rate and spectator relaxation rate, see Eq.~\eqref{eqn:2q_tr}, accurately predicts the effect of SDID (solid red line).The only fitting parameter in our model is a simple multiplicative rescaling parameter to adjust for state-preparation and measurement error (SPAM), whereby we multiply the theory prediction by the value of the initial ($\tau=0$) experimental data point. Heuristically, we find that the SDID-enhanced dephasing is well approximated by an exponential with a rate proportional to $1/T_2^{(0)}+1/T_1^{(1)}$. This is shown in Fig.~\ref{fig1}a), where the black dashed line is a single exponential decay curve with the enhanced dephasing rate and the overall scaling to adjust for SPAM. As we will discuss later, this simple heuristic only holds in the case of $\nu \gg 1/T_1^{(1)}$. However, for our particular case $\nu \approx 1/T_1^{(1)}$, which explains the oscillations observed in the curves for the full theoretical model in Fig.~\ref{fig1}.

In Fig.~\ref{fig1}b) we show that SDID is an additive effect (in terms of the dephasing rate) when additional spectators are involved. Here, we have a central control qubit coupled to three spectator qubits, $Q_{1,2,3}$, with $ZZ$-coupling strengths $4\nu_j=\{\SI{47}{\kilo\hertz},\SI{48}{\kilo\hertz},\SI{41}{\kilo\hertz}\}$, spectator lifetimes $T_1^{(j)}=\{\SI{150}{\micro\second},\SI{218}{\micro\second},\SI{122}{\micro\second}\}$ and control qubit $T^{(0)}_2=\SI{241}{\micro\second}$. One can clearly see that with every additional spectator qubit being prepared in the $\ket{1}$-state, starting from spectator $Q_1$, the dephasing rate of the control qubit increases. Our multi-qubit theory model, see Eq.~\eqref{eqn:Nq_tr}, accurately captures the control qubit dephasing for an increasing number of spectators (colored solid lines). As in Fig.~\ref{fig1}a), to account for SPAM we rescale each theory curve by the value of the first data point of the corresponding experimental data.

The starting point of our theoretical analysis of SDID is the master equation describing the evolution of a $ZZ$-coupled array of qubits with relaxation and dephasing
\begin{align}
    &\dot\rho = -i\left[\hat{H},\rho\right] + \sum_{j=0}^{N}\left(\gamma_j\mathcal{D}\left[\hat{\sigma}_j^-\right]\rho + \gamma^\phi_j\mathcal{D}\left[\hat{Z}_j\right]\rho \right), \label{eqn:me_model} \\
    &\hat{H} = \sum_{j=1}^N \nu_j \hat{Z}_0 \otimes \hat{Z}_j, \label{eqn:ZZHam}
\end{align}
where $\mathcal{D}[\hat{x}]\rho = \hat{x}\rho\hat{x}^\dagger - \left\{\hat{x}^\dagger\hat{x},\rho\right\}/2$. Note that $\hat{H}$ is in the rotating frame with respect to the qubits' self-Hamiltonian $\hat{H}_{\rm self} = \sum_j (\omega_j/2)\hat{Z}_j$. $\gamma_j = 1/T_1^{(j)}$ and $\gamma_j^\phi$ are the relaxation and pure dephasing rates of qubit $j$, and $\nu_j$ is the $ZZ$-coupling strength between spectator qubit $j$ and control qubit $0$. For analysis of the dynamics of the control qubit we can neglect $ZZ$-coupling between the spectators.

In Appendix \ref{app:ME} we discuss the microscopic derivation of Eq.~\eqref{eqn:me_model}, but here we focus on obtaining a model for the evolution of the coherence of the reduced state of the control qubit. To that end, we can express $\rho(t)$ as
\begin{align}
    \nonumber\rho(t) &= \ketbra{0}\otimes\rho_{00}(t) + \ketbra{1}\otimes\rho_{11}(t) \\ &+ \ketbra{0}{1}\otimes\rho_{01}(t) + \ketbra{1}{0}\otimes\rho_{10}(t), \label{eqn:rho_decomp}
\end{align}
where the spectator qubit operators $\rho_{jk}(t)$ contain all time-dependence of $\rho(t)$. Note that while $\rho_{00}$ and $\rho_{11}$ are valid density matrices, $\rho_{01}$ is not required to be, but we have that $\rho_{10} = \rho_{01}^\dagger$. Similar to how measurement-induced dephasing is analysed in circuit QED \cite{Gambetta06}, the trace of $\rho_{01}(t)$ captures the evolution of the coherence of the control-qubit reduced state.

The operator $\rho_{01}$ evolves under the equation
\begin{align}
    \nonumber\dot\rho_{01} &= \left(\bra{0}\otimes\iden\right)\dot\rho\left(\ket{1}\otimes\iden\right) = -\tilde\gamma_0\rho_{01} + i\sum_{j=1}^N \nu_j\left\{\hat{Z}_j,\rho_{01}\right\} \\
    &+\sum_{j=1}^{N}\left(\gamma_j\mathcal{D}\left[\hat{\sigma}_j^-\right]\rho_{01} + \gamma_j^\phi\mathcal{D}\left[\hat{Z}_j\right]\rho_{01}\right),
\end{align}
where $\tilde\gamma_0 = \gamma_0^\phi + \gamma_0/2 = 1/T_2^{(0)}$ characterizes the dephasing of the control qubit due to its intrinsic decoherence. We can immediately remove the dynamics due to $\tilde\gamma_0$ by defining $\eta = e^{\tilde\gamma_0t}\rho_{01}$, which evolves as
\begin{align}
    \dot\eta = \sum_{j=1}^N \left(i\nu_j\left\{\hat{Z}_j,\eta\right\} + \gamma_j\mathcal{D}\left[\hat{\sigma}_j^-\right]\eta + \gamma_j^\phi\mathcal{D}\left[\hat{Z}_j\right]\eta\right). \label{eqn:01dyn}
\end{align}
As the structure of this equation shows, the part of $\eta$ corresponding to each spectator evolves independently of the others, so that if the system starts in a tensor product structure it will remain in one, i.e.~$\eta(t) = \bigotimes^N_{j=1}\eta_j(t)$. Thus, solving for one spectator, we can simply iterate and combine by tensor product to solve for all spectators.

To calculate the trace of $\eta$, we need only solve for its diagonal elements. For each spectator, we find that
\begin{align}
    &[\eta_j(t)]_{00} = [\eta_j(0)]_{00}e^{-2i\nu_j t} \\
    &[\eta_j(t)]_{11} = \frac{[\eta_j(0)]_{11}}{4i\nu_j - \gamma_1}\left( 4i\nu_j e^{\left(2i\nu_j  - \gamma_1\right)t}  - \gamma_1e^{-2i\nu_j t}\right)
\end{align}
where $[\eta_j(t)]_{kk}$ is the diagonal element corresponding to the spectator in state $\ketbra{k}$. Using this, for a single spectator ($N=1$, $\nu \equiv \nu_1$) we can obtain the analytical expression for the trace of $\rho_{01}(t)$
\begin{align}
   \nonumber &{\rm Tr}\left[\rho_{01}(t)\right] = [\rho_{01}(0)]_{00} e^{-\left(2i\nu  + \tilde\gamma_0\right)t}  \\ &+ [\rho_{01}(0)]_{11} \frac{e^{-\tilde\gamma_0t}}{4i\nu - \gamma_1}\left( 4i\nu e^{\left(2i\nu  - \gamma_1\right)t}  - \gamma_1e^{-2i\nu t}\right), \label{eqn:2q_tr}
\end{align}
which has a clear interpretation. If the spectator is in the ground state initially, then the control qubit evolution is given by the first line of Eq.~\eqref{eqn:2q_tr}, which is just the usual coherent evolution driven by the $ZZ$-coupling and the control-qubit intrinsic dephasing. For an excited spectator, the decay envelope is given by the second line of Eq.~\eqref{eqn:2q_tr}. While more complicated, in the limit $\nu \gg 1/T_1^{(1)}$ the absolute value of this simplifies to exponential decay at the enhanced rate $1/T_2^{(0)} + 1/T_1^{(1)}$. 

For multiple spectators that each start in a Pauli-$Z$ eigenstate, the trace of $\rho_{01}(t)$ is given by
\begin{align}
    &{\rm Tr}\left[\rho_{01}(t)\right] = \\ \nonumber& e^{-\tilde\gamma_0t}\prod_{s_j = 0}^N e^{-2i\nu_j t}\prod_{s_j = 1}^N \frac{\left( 4i\nu_j e^{\left(2i\nu_j  - \gamma_j\right)t}  - \gamma_je^{-2i\nu_j t}\right)}{4i\nu_j - \gamma_j}, \label{eqn:Nq_tr}
\end{align}
where as before $s_j$ labels the initial state of the $j$'th spectator qubit. As was the case for a single qubit, in the limit $\nu_j \gg 1/T_1^{(j)}$ the control qubit coherence envelope simplifies to exponential decay at the enhanced rate $1/T_2^{(0)} + \sum_{j, s_j = 1} 1/T_1^{(j)}$, which agrees with what we observe in experiment.

The theory model we have so far considered does not include the Hahn-echo applied in our experimental protocol. However, as we will explain in more detail in section \ref{sec:cpmg}, when echo sequences are applied the important consideration is the maximum possible phase-accumulation of the control-qubit due to its interaction with the spectator qubits in the $\ket{1}$-state. For Ramsey decay with no echo, for a total evolution time of $\tau$ the accumulated phase is contained within $\left[0,2\nu \tau\right]$, with the lower bound when the spectator relaxes exactly at $t=\tau/2$, and the upper bound when the spectator does not relax. For Ramsey decay with a single echo, the accumulated phase is also contained within $\left[0,2\nu \tau\right]$, but with the lower bound for no spectator relaxation, and the upper bound when the spectator relaxes at $t=\tau/2$. Thus, the model for Ramsey without echo accurately describes the situation of Ramsey with echo, as evidenced by the excellent agreement between theory and experiment in Fig.~\ref{fig1}.

\section{Composite Pulse Sequences}
\label{sec:cpmg}

\begin{figure}[t!]
    \centering
    \includegraphics[width=0.48\textwidth]{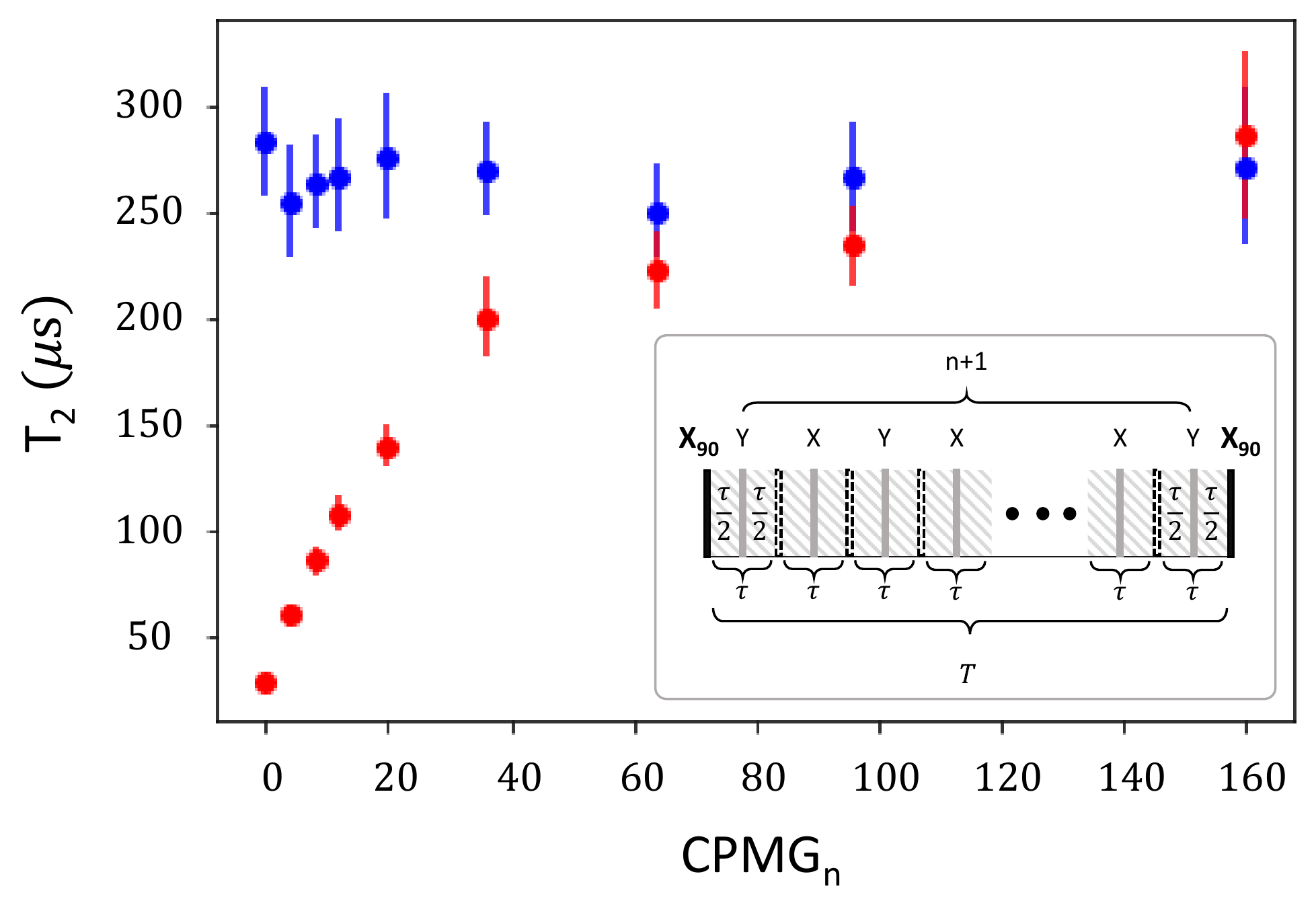}
    \caption{Dephasing time $T_2$ is measured for a control qubit as a function of increasing CPMG-sequence order. The control is $ZZ$-coupled to three spectator qubits, and all spectators are prepared in the $\ket{0}$-state (blue) or in the $\ket{1}$-state (red). Inset: Experimental sequence with CPMG pulses of $n^\mathrm{th}$-order inserted between the Ramsey pulses Dashed lines are guides for the eyes.}
    \label{fig2}
\end{figure}

The nature of the SDID effect can be intuitively understood as a single phase-kick due to the $ZZ$-coupling when an excited spectator qubit relaxes. The control qubit's frequency depends on the spectator qubit's state with a frequency difference of $4\nu_j$, and the control qubit frequency for all the experiments presented here is calibrated while the spectator(s) are in the ground state. The single Hahn-echo previously described cancels any static frequency shift due to the spectator(s) being in the excited state as long as they stay in this state. While a single spectator relaxation results in a coherent phase evolution for each quantum trajectory, the random nature of $T_1$-relaxation leads to dephasing in the ensemble average. 

If one could predict the exact moment when a $T_1$ event occurs, SDID could be fully reversible. Since this is impossible, the closest we can do is to implement a decoupling sequence such as CPMG \cite{Carr1954, Meiboom1958}. With increasing order, this can refocus more of the SDID-related phase kicks. In Fig.~\ref{fig2} we measure the dephasing rate of the control qubit with a Ramsey experiment as a function of increasing CPMG$_n$ pulse sequence order, where $n+1$ $Y/X$-pulses are inserted with temporal symmetry and equidistant spacing between the two Ramsey pulses, see inset Fig. \ref{fig2}. Here, we take the same set of four qubits as presented in Fig.~\ref{fig1} b). Without any CPMG pulses the control $T_2^{(0)}\approx\SI{280}{\micro\second}$ drops to $T_2^{(0)}\approx\SI{30}{\micro\second}$ when all three spectators are in state $\ket{1}$. The data presented in Fig.~\ref{fig2} shows a clear revival of $T_2^{(0)}$ as we increase the CPMG$_n$-order, with a full revival at $n\approx160$.

To model reduction of the SDID effect due to CPMG$_n$, the relevant quantities are the lower and upper bounds on the amount of phase that the control qubit can accumulate due to the $ZZ$-interaction with the spectator qubits. As shown in the inset of Fig.~\ref{fig2}, we can divide the total time $T$ of a CPMG$_n$ experiment into periods of length $\tau = T/(n+1)$, where each period has an $X$- or $Y$-pulse in the middle. If the sign of the $ZZ$-interaction remains constant during a $\tau$-period, the total phase accumulation will be zero due to the echoing effect of the $X/Y$-pulse. Thus, to understand the total random phase accumulation during the full time evolution, we need only consider the $\tau$-periods when a spectator relaxation occurs.

For a single spectator, there is only one such period, and the phase accumulation is lower bounded by $0$ (spectator relaxation at the start or end of the $\tau$-period), and upper bounded by $2\nu\tau = 2\nu T/(n+1)$ (spectator relaxation in the middle of the $\tau$-period). For $n=0$, the sequence is equivalent to the Hahn-echo sequence of Fig.~\ref{fig1}, and we obtain that the accumulated phase is contained within $\left[0,2\nu T\right]$, exactly as would be the case if there were no Hahn-echo. This explains why the theoretical model with no Hahn-echo accurately predicts the results of the sequence with Hahn-echo. For any CPMG$_n$ we can define an effective $\nu' = \nu/(n+1)$, and use $\nu'$ instead of $\nu$ in our theory model for Ramsey decay. The generalization to multi-spectators is straightforward, as the phase evolution due to each spectator can be considered independently, and $\nu'_j$ replaces $\nu_j$ in Eq.~\eqref{eqn:Nq_tr}.

\section{Randomized Benchmarking}
\label{sec:RB}

\begin{figure}
    \centering
    \includegraphics[width=0.48\textwidth]{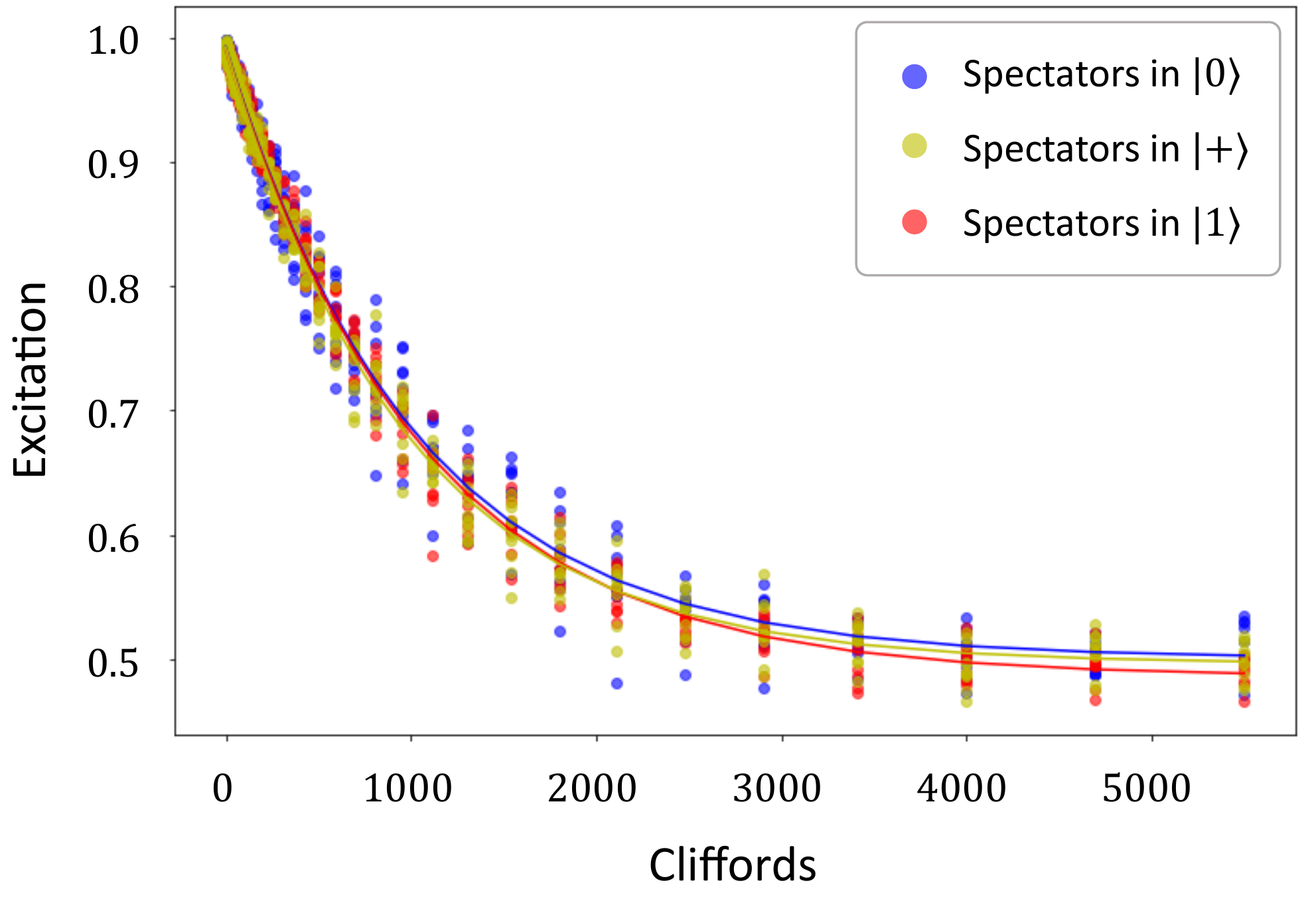}
    \caption{Single qubit randomized benchmarking on a control qubit coupled to three spectator qubits.  Dots are experimental data and solid lines are exponential fits to data. From the fitting we get an error per Clifford of $9.47(14)\times10^{-4}$ with all spectators in $\ket{0}$ (blue), $10.1(1)\times10^{-4}$ with spectators in $\ket{+}$ (yellow) and $9.44(9)\times10^{-4}$ with spectators in $\ket{1}$ (red).}
    \label{fig3}
\end{figure}

To understand the impact of SDID on RB we calculate the quantum channels describing the evolution of the control qubit conditioned on the initial and final state of the spectator qubit. To start, we consider the full two-qubit channel generated by Eq.~\eqref{eqn:me_model} for a single spectator and no control-qubit dissipation. This is most easily expressed in vectorized form as
\begin{align}
    \superket{\rho(t)} = e^{\mathcal{L}t}\superket{\rho(0)},
\end{align}
where $\superket{\rho}$ is a vector formed by column-stacking the density matrix $\rho$, and $e^{\mathcal{L}t}$ is the superoperator description of the two-qubit channel, with $\mathcal{L}$ the associated Liouvillian superoperator of Eq.~\eqref{eqn:me_model}. To define the conditional control-qubit channels $\hat{\Lambda}_{ij}$ for the spectator initial state $\ket{i}$ and final state $\ket{j}$, we start in the initial state $\rho_i = \rho_c\otimes\ketbra{i}$, and project onto the final spectator state $\ketbra{j}$ via the operation
\begin{align}
    \hat{\Lambda}_{ij}\superket{\rho_c} = \frac{1}{\mathcal{N}_{ij}} \left(\iden\otimes\bra{j}\otimes\iden\otimes\bra{j}\right)e^{\mathcal{L}t}\superket{\rho_i}. 
\end{align}
Here $\mathcal{N}_{ij}$ is a normalization factor given by the trace of $\hat{\Lambda}_{ij}\superket{\rho_c}$, i.e.~the probability that the spectator starts in state $\ket{i}$ and ends in state $\ket{j}$, which in vectorized notation is defined as
\begin{align}
    \mathcal{N}_{ij} =  \sum_{n=0}^1\left(\iden\otimes\bra{n}\otimes\iden\otimes\bra{n}\right)\hat{\Lambda}_i\superket{\rho_c} . 
\end{align}

Since the spectator qubit cannot be excited by the dynamics we consider, the process described by $\hat{\Lambda}_{01}$ cannot occur, and so it is the all zero matrix. The other $\hat{\Lambda}_{ij}$ are diagonal matrices with structure
\begin{align}
    {\rm diag}[\hat{\Lambda}_{ij}] = \left(1,\lambda_{ij},\lambda^*_{ij},1\right).
\end{align}
where $\lambda^*$ is the complex conjugate of $\lambda$. In detail, we have that
\begin{align}
    &\lambda_{00} = \lambda^*_{11} = e^{i2\nu t}, \\
    &\lambda_{10} = \frac{\gamma_1\left(e^{i2\nu t} - e^{-\left(\gamma_1 + i2\nu\right) t} \right)}{\left(\gamma_1+i4\nu\right)\left(1-e^{-\gamma_1 t}\right)}.
\end{align}

Without SPAM, a single-qubit RB curve has the form $p^n + (1-p^n)/2$ after $n$ gates, and the RB decay constant, $p$, for a single-qubit error process $\Lambda$ is related to the trace of its superoperator via the expression \cite{Emerson_2005}
\begin{align}
    p = \frac{{\rm Tr}[\hat{\Lambda}] - 1}{3}.
\end{align}
Thus, we have for each of our conditional processes that
\begin{align}
    p_{ij} = \frac{1 + 2{\rm Re}[\lambda_{ij}]}{3},
\end{align}
if the error process $\hat{\Lambda}_{ij}$ were to happen for every gate in the RB sequence. For $\hat{\Lambda}_{00}$ and $\hat{\Lambda}_{11}$ this is possible, as repeated action of these error processes represents, respectively, the evolution with a ground state spectator, and when the excited-state spectator does not relax. In both cases we find the same RB decay constant
\begin{align}
    p_{00} = p_{11} = \frac{1}{3}\left(1 + 2\cos(2\nu t)\right).
\end{align}
This means that RB is insensitive to the sign of the $ZZ$-interaction, which is a consequence of the twirling of the error channel implemented by the random Clifford gates.

For the situation when the spectator does relax, the evaluation is not so simple, as the spectator can only relax once. Assuming that the relaxation occurs during the $m$'th gate, then the first $m-1$ gates experience the error channel $\hat{\Lambda}_{11}$, and the last $n-m$ the error channel $\hat{\Lambda}_{00}$ as the spectator now starts in the ground state. Only the $m$'th gate experiences the channel $\hat{\Lambda}_{10}$. As $\hat{\Lambda}_{11}$ and $\hat{\Lambda}_{00}$ lead to the same RB decay constant, a single action of $\hat{\Lambda}_{10}$ will have an inconsequential impact on the RB decay curve. 

The coherent error induced by the $ZZ$-interaction found in $\hat{\Lambda}_{11}$ and $\hat{\Lambda}_{00}$ plays a dominant role, and as RB is insensitive to the sign of this interaction, the spectator relaxation cannot refocus any of this coherent error except during the gate when the relaxation occurs. In the limit of long RB sequences, this causes only a small deviation from the exponential decay predicted with no spectator relaxation. As such, RB is effectively insensitive to the effects of SDID.

The above expressions for RB decay assume that the single-qubit Clifford gates on the control qubit are calibrated with a drive frequency corresponding to the ``bare'' control qubit frequency.  For such a calibration, in the frame rotating at the drive frequency there is an equal magnitude and opposite sign $ZZ$-interaction for both spectator qubit states. In our experiments, the gates are calibrated with the spectator qubits in the ground state, and the drive frequency corresponds to the bare control qubit frequency shifted by the $ZZ$-interaction for ground-state spectators.

A simple modification of our RB theory model is required to account for the experimental calibration. For a ground-state spectator, the control qubit experiences no net shift due to the $ZZ$-interaction, while for an excited-state spectator it experiences twice the $ZZ$-interaction of the bare frequency calibration. Thus, it is straightforward to see that the RB decay constants for the experimental calibration are
\begin{align}
    p^{\rm exp}_{00} = 1,~p^{\rm exp}_{11} = \frac{1}{3}\left(1 + 2\cos(4\nu t)\right).
\end{align}
Nevertheless, RB remains insensitive to the effects of spectator relaxation.

In Fig.~\ref{fig3} we show experimental RB curves for the three spectator qubits initially all prepared in the $\ket{0}$, $\ket{1}$, or $\ket{+} \propto \ket{0} + \ket{1}$ state. The decay curves are nearly indistinguishable, and the error per Clifford (EPC) estimated from each differs at the $10^{-5}$ level. As expected, to this level of estimation accuracy we see no impact of SDID on RB. While our Ramsey experiments strikingly demonstrated the impact of SDID, no such behavior is observed for the RB decay.

However, given our theory model, for the experimental calibration we would expect $\ket{0}$ to have a smaller EPC than the other states, as it should have no coherent error due to the $ZZ$-interaction, while the $\ket{1}$ and $\ket{+}$ states should. Quantitatively, for the device parameters of our system, the upper bound on the EPC due to the $ZZ$-interaction (if all $\nu_i$ add constructively) is on the order of $10^{-4}$. Thus, is is likely that the RB decay is also strongly impacted by some other source of error not contained in our model.


\section{Conclusion}
\label{sec:Conc}

In this manuscript we have experimentally investigated the effective dephasing induced on a qubit due to the spontaneous decay of spectator qubits coupled via a $ZZ$-interaction. Our experimental results are supported by a robust theoretical model, and we find excellent agreement between theory and experiment with only a single fit parameter to account for state preparation and measurement error. We have shown that the SDID mechanism can be the dominate source of decay in a Ramsey experiment, but owing to the shot-by-shot coherent nature of the noise, it can be corrected for using dynamical decoupling sequences such as CPMG. Furthermore, we have found both in theory and experiment that randomized benchmarking is insensitive to SDID.

The quantitative impact of SDID on specific quantum information protocols remains an open question for future study. The SDID error mechanism is intrinsically correlated, as a decay event on one qubit induces ensemble average dephasing on another. This may be particularly damaging to applications known or expected to be sensitive to correlated errors, such as quantum error correction, entanglement generation, or quantum metrology. Moreover, protocols with long qubit idle times, such as quantum memories, are likely susceptible to the impact of SDID. On the other hand, we have shown that SDID can be mitigated using high-order dynamical decoupling, and random circuits seem to have some degree of intrinsic insensitivity to SDID, as evidenced by our results for randomized benchmarking. 

SDID is one example of an error that emerges due to the multi-qubit interplay of single-qubit error mechanisms, where the effect is ``worse than the sum of its parts''. With quantum information protocols continuing to grow in size, duration, and complexity, we expect more of these emergent errors to appear. This highlights the importance of continued experimental and theoretical research into the impact of simultaneous error mechanisms in combination, not in isolation.

\acknowledgements

We thank Seth Merkel for discussions of the randomized benchmarking model, as well as David McKay, Haggai Landa, and Oliver Dial for careful reading of the draft. This research was sponsored by the Army Research Office and was accomplished under Grant Number W911NF-21-1-0002. The views and conclusions contained in this document are those of the authors and should not be interpreted as representing the official policies, either expressed or implied, of the Army Research Office or the U.S.~Government. The U.S.~Government is authorized to reproduce and distribute reprints for Government purposes notwithstanding any copyright notation herein.

\appendix

\section{Device Parameters}
\label{app:device}

\begin{table}[!h]
    \centering
    \begin{tabular}{|l|c|c|c|}
    \hline
         Qubit & $T_1$ ($\mu$s) & $T_2$ ($\mu$s) & $4\nu_j$ (kHz) \\
         \hline
         \multicolumn{4}{l}{\bf Fig.~\ref{fig1}a)} \\
         \hline
         Control $Q_0$ & - & 127 & -\\
         \hline
         Ancilla $Q_1$& 107 & - & 45\\
         \hline
         \multicolumn{4}{l}{\bf Figs.~\ref{fig1}b), \ref{fig2}, and \ref{fig3}} \\
         \hline
         Control $Q_0$& 141 & 241 & -\\
         \hline
         Ancilla $Q_1$ & 150 & 258 & 47\\
         \hline
         Ancilla $Q_2$ & 218 & 400 & 48\\
         \hline
         Ancilla $Q_3$ & 122 & 175 & 41 \\
         \hline
    \end{tabular}
    \caption{Summary of the relevant parameters of the qubits from the IBM Quantum Falcon R5 device used in this manuscript. Parameters for Fig.~\ref{fig1}a) were measured during our experiment (control $T_1$ and spectator $T_2$ were not measured as they are not needed), while parameters for Figs.~\ref{fig1}b), \ref{fig2}, and \ref{fig3} were taken from indpendent device calibration.}
    \label{tab:devic}
\end{table}

\section{Derivation of the Lindblad-form master equation}
\label{app:ME}

In this appendix, we will microscopically derive the Lindblad-form master equation of Eq.~\eqref{eqn:me_model}, and justify its use as the starting point for our theoretical model. To understand why this is necessary, notice that the Lindbladian evolution term
\begin{align}
    \dot\rho = \mathcal{L}[\rho] =  \sum_{j=0}^{N}\left(\gamma_j\mathcal{D}\left[\hat{\sigma}_j^-\right]\rho + \gamma^\phi_j\mathcal{D}\left[\hat{Z}_j\right]\rho \right), \label{eqn:targetME}
\end{align}
has a time-dependent terms in the interaction frame defined by the system Hamiltonian
\begin{align}
    \hat{H} = \sum_{j=0}^N\frac{\omega_j}{2} \hat{Z}_j + \sum_{j=1}^N\nu_j \hat{Z}_0 \hat{Z}_j, \label{eqn:logHam}
\end{align}
due to the $ZZ$-coupling. Thus, $\mathcal{L}$ cannot be microscopically derived using the standard Born-Markov rotating-wave-approximation (BMRWA) approach \cite{Breuer07,Lidar19} which precludes a time-dependent Lindbladian, as the strong secular approximation made in the BMRWA removes all interference between decay pathways that have non-degenerate Bohr frequencies (energy differences between the eigenstates of $\hat{H}$).

Without a clear microscopic derivation, we must treat Eq.~\eqref{eqn:me_model} as a phenomenological model. Beyond justifying its use as the starting point of our analysis, microscopic derivation enables us to clarify the regime of validity of Eq.~\eqref{eqn:me_model}, and check for missing elements of the model that may be physically relevant. Given the excellent agreement we see with experimental results, we do not expect Eq.~\eqref{eqn:me_model} to be missing anything in the parameter regime of our experiments. As we will soon see, this is indeed the case, but we will also comment on what would happen outside this regime, such as with stronger $ZZ$-coupling.

The starting point of our analysis is the lab-frame description of an array of transversely coupled superconducting qubits
\begin{align}
    \hat{H}_{0}= \sum_{j=0}^N \frac{\omega_j}{2} \hat{Z}_j + \sum_{k>j}^Ng_{jk} \hat{X}_j  \hat{X}_k. \label{eqn:physHam}
\end{align}
We assume that each of these qubits couple to independent electromagnetic baths via the system-bath interaction Hamiltonian 
\begin{align}
    \hat{H}_{{\rm SB}, 0} = \sum_{j=0}^N \tilde{\xi}_j \hat{X}_j \otimes \left(\hat{B}_j + \hat{B}_j^\dagger\right),
\end{align}
where $\hat{B}_j$ is the SB-coupling operator for the bath of qubit $j$, with $\tilde{\xi}$ the interaction strength. The assumption of independent baths is not necessary for this derivation, but simplifies the analysis. Note that given our choice of $\hat{H}_{{\rm SB}, 0}$ we are focusing solely on the derivation of the energy relaxation part of Eq.~\eqref{eqn:targetME}, as this is the part that does not commute with the Hamiltonian of Eq.~\eqref{eqn:logHam} and leads to a time-dependent Lindbladian.

Equation \eqref{eqn:physHam} defines a Hamiltonian $\hat{H}_0$ composed of what are usually referred to as the \emph{physical} qubits. Diagonalization of $\hat{H}_0$ results in a Hamiltonian of the form  \cite{Berke20}
\begin{align}
      \nonumber\hat{H}_{L} &= \sum_{j=0}^N \frac{\omega_j}{2} \hat{Z}^L_j + \sum_{k>j}^N \nu_{jk} \hat{Z}^L_j  \hat{Z}^L_k \\ &+ \sum_{n>k>j}^N \mu_{jkn} \hat{Z}^L_j  \hat{Z}^L_k \hat{Z}^L_n + ...
\end{align}
that describes an array of \emph{logical} qubits coupled by diagonal interactions consisting of many-body $Z$-terms. The logical qubits incorporate the hybridization of the physical qubits due to their transverse coupling, and $\hat{H}$ of Eq.~\eqref{eqn:logHam} is the truncation of the above expression to two-body interactions. This truncation is accurate for weak enough transverse coupling \cite{Berke20}, and is the first approximation in deriving Eq.~\eqref{eqn:me_model}.

The system-bath interaction transformed to the basis of logical qubits is generally complicated. The physical qubit operator $\hat{X}_j$ is transformed to a logical qubit operator whose dominant component is $\hat{X}^L_j$, but includes small corrections due to the qubit hybridization. To simply the analysis, we focus on the situation described by Eq.~\eqref{eqn:logHam}, where there is no $ZZ$-coupling between the spectator qubits.

In this case, the system-bath Hamiltonian for the logical qubits including the lowest order correction is
\begin{align}
   \nonumber &\hat{H}_{{\rm SB}, L} = \xi_0\left(\hat{X}^L_0 + \sum_{j=1}^Na_{0j}\hat{Z}^L_0 \hat{X}^L_j\right) \otimes \left(\hat{B}_0 + \hat{B}_0^\dagger\right)\\
    &+\sum_{j=1}^N \xi_j \left(\hat{X}^L_j + a_{j0}\hat{X}^L_0 \hat{Z}^L_j\right) \otimes \left(\hat{B}_j + \hat{B}_j^\dagger\right), \label{eqn:logHSB}
\end{align}
where $a_{0j}$ and $a_{j0}$ are mixing coefficients describing the hybridization, and $\xi_j$ is a renormalized system-bath interaction strength that makes the coefficient of the dominant $\hat{X}^L_j$ term unity. We are typically in a parameter regime where $\abs{a_{0j}}, \abs{a_{j0}} \ll 1$, and so have dropped terms  $\mathcal{O}\left(\abs{a_{0j}}^2\right)$ and higher that hybridize the spectator qubits through an interaction mediated by the control qubit.

As can be seen, the dominate correction in the logical qubit basis is to introduce a decoherence channel for qubit $j$ through the control qubit's bath, and vice versa. Using this system-bath interaction we will now attempt three microscopic derivations of Eq.~\eqref{eqn:me_model}: the first, following the BMRWA approach, will fail, which will motivate us to consider a second, ultimately successful, approach using a partial secular approximation (BMPSA) \cite{Trushechkin21}. We will also employ a third approach, a cumulant expansion and time coarse-graining (CETCG) \cite{Lidar19} to understand the intermediate regime between our first two approaches. From here onward we will suppress the superscript $L$, as all equations are in the logical qubit basis.

\subsection{Born-Markov rotating-wave-approximation: a failed derivation}

In the BMRWA approach, the derivation of the Lindbladian terms for each independent bath can be treated separately. As such, for illustrative purposes we can consider a two-qubit system and derive the master equation for the spectator's interaction with its bath, as described by the full Hamiltonian
\begin{align}
    &\nonumber \hat{H}_{2{\rm Q}} = \hat{H}_{\rm S} + \hat{H}_{\rm B} + \hat{H}_{\rm SB}, \\
    &\nonumber\hat{H}_{\rm S} = \frac{\omega_0}{2} \hat{Z}_0 + \frac{\omega_1}{2} \hat{Z}_1 + \nu \hat{Z}_0 \hat{Z}_1, \\ 
    &\hat{H}_{\rm SB} = \xi \left(\hat{X}_1 + a\hat{X}_0\hat{Z}_1\right) \otimes \left(\hat{B}_j + \hat{B}_j^\dagger\right), \label{eqn:fullham2Q}
\end{align}
where $\hat{H}_{\rm B}$ is the bath Hamiltonian. As will become clear during the derivation, using the Linbladian we obtain from Eq.~\eqref{eqn:fullham2Q} we will be able to determine the full Linbladian for an arbitrary number of spectator qubits, including the contribution due to the control qubits interaction with its bath.

The first step is to go to the interaction frame with respect to $\hat{H}_{\rm S}$ and $\hat{H}_{\rm B}$, which results in the Hamiltonian
\begin{align}
    &\nonumber\hat{H}_{2{\rm Q}}' = \xi \Big(e^{-i\omega_1 t}\left[e^{-i2\nu t}\ketbra{1} + e^{i2\nu t}\ketbra{0} \right]\otimes\hat{\sigma}^- \\
    &\nonumber+ ae^{-i\omega_0 t}\hat{\sigma}^-\otimes\left[e^{-i2\nu t}\ketbra{1} - e^{i2\nu t}\ketbra{0}\right] + h.c. \Big)\\
    &\otimes \left(\hat{B}_1(t) + \hat{B}_1^\dagger(t)\right) \label{eqn:Hinter}
\end{align}
where we have separated terms in the system operator by their oscillation frequency. The rest of the derivation follows the standard procedure of the BMRWA approach, as outlined in many places, for example Refs.~\cite{Breuer07,Lidar19}. Crucially, this approach makes a \emph{strict} rotating-wave-approximation to drop all explicitly time-dependent terms in the Lindbladian expressed in the interaction frame. 

Equivalently, the RWA step can be thought of as writing the system operator of $\hat{H}_{\rm SB}$ in the eigenbasis of $\hat{H}_{\rm S}$, and separating the resulting terms into groups determined by their Bohr frequency, i.e.~the eigenenergy difference of the eigenstates connected by each term of $\hat{H}_{\rm SB}$. Each Bohr frequency group then results in a separate Lindblad operator in the (diagonal) Lindbladian. Explicitly, for a system operator $\hat{X}$ in $\hat{H}_{\rm SB}$, and eigenvalues/eigenvectors $\lambda_k/\ket{\lambda_k}$ of $\hat{H}_{\rm S}$, the BMRWA master equation is
\begin{align}
    \dot\rho = \sum_{\Omega}\Gamma[\Omega]\mathcal{D}[\hat{L}_\Omega]\rho, \label{eqn:BMME}
\end{align}
where $\Omega$ are the Bohr frequencies, with corresponding Lindblad operators
\begin{align}
    \hat{L}_\Omega = \sum_{\lambda_j - \lambda_k = \Omega}\bra{\lambda_k}\hat{X}\ket{\lambda_j}\ketbra{\lambda_k}{\lambda_j}, \label{eqn:LinOps}
\end{align}
and the decay rate $\Gamma(\Omega)$ is a function of the bath and will be defined shortly.

For our system described by Eq.~\eqref{eqn:Hinter} the eight Bohr frequencies are $\pm\omega_0 \pm 2\nu$ and $\pm\omega_1 \pm 2\nu$, which for $\omega_{0/1} \gg \abs{\nu}$ and well detuned qubits are all non-degenerate. The operators corresponding to each of these Bohr frequencies (written in the eigenbasis of $\hat{H}_{\rm S}$), are exactly the terms oscillating at the respective Bohr frequency in Eq.~\eqref{eqn:Hinter}. Thus, we can write down the BMRWA master equation
\begin{align}
    \dot{\rho} =~&\Gamma\left(\omega_1 + 2\nu\right)\mathcal{D}\left[\ketbra{1}\otimes\hat{\sigma}^-\right]\rho \\ \nonumber&+ \Gamma\left(\omega_1 - 2\nu\right)\mathcal{D}\left[\ketbra{0}\otimes\hat{\sigma}^-\right]\rho \\
    \nonumber&+  \abs{a}^2\Gamma\left(\omega_0 + 2\nu\right)\mathcal{D}\left[\hat{\sigma}^-\otimes\ketbra{1}\right]\rho \\
    &+ \abs{a}^2\nonumber\Gamma\left(\omega_0 - 2\nu\right)\mathcal{D}\left[\hat{\sigma}^-\otimes\ketbra{0}\right] \rho,
\end{align}
where we have assumed a zero-temperature bosonic bath such that the terms corresponding to qubit excitation vanish as the bath has no spectral weight at Bohr frequencies $\approx-\omega_{0/1}$. The decay rate for each Lindblad operator is a function of the Bohr frequency given by
\begin{align}
    \Gamma(\Omega) = \abs{\xi}^2\int_{-\infty}^{\infty} dt~e^{i\Omega t} \left<\hat{B}_1^\dagger(t)\hat{B}_1(0)\right>_{\rho_{\rm B}}. \label{eqn:decayrate}
\end{align}
where the expectation value is taken with respect to the stationary bath state $\rho_{\rm B}$, which for a zero-temperature bosonic environment is the multi-mode vacuum. Note that we have not explicitly included the Lamb-shift Hamiltonian term arising from the system-bath interaction, as it only amounts to a weak renormalization of the existing system eigenenergies.

As we have previously dropped terms of $\mathcal{O}(\abs{a}^2)$ we do so again to arrive at the final BMWRA master equation
\begin{align}
    \dot{\rho} = \Gamma_+\mathcal{D}\left[\ketbra{1}\otimes\hat{\sigma}^-\right]\rho + \Gamma_-\mathcal{D}\left[\ketbra{0}\otimes\hat{\sigma}^-\right]\rho, \label{eqn:BMRWA_ME}
\end{align}
where we use $\Gamma_{\pm} = \Gamma\left(\omega_1 \pm 2\nu\right)$ as shorthand notation. Clearly, this is not the same as the relevant part of Eq.~\eqref{eqn:targetME} that we are aiming for
\begin{align}
    \dot{\rho} = \gamma_1\mathcal{D}\left[\hat{\mathbb{I}}\otimes\hat{\sigma}^-\right]\rho = \gamma_1\mathcal{D}\left[\left(\ketbra{1} + \ketbra{0}\right)\otimes\hat{\sigma}^-\right]\rho,
\end{align}
where there is a single rate $\gamma_1$, and the Lindblad operators $\ketbra{j}\otimes\hat{\sigma}^-$ add coherently, as opposed to incoherently as they do in Eq.~\eqref{eqn:BMRWA_ME}. If we assume $\Gamma_+ \approx \Gamma_- \equiv \Gamma_1$, which is valid for a bath spectral density that around $\omega_1$ varies little if the frequency changes by $\nu \ll \omega_1$, then we can express Eq.~\eqref{eqn:BMRWA_ME} as 
\begin{align}
    \dot{\rho} = \frac{\Gamma_1}{2}\mathcal{D}\left[\hat{\mathbb{I}}\otimes\hat{\sigma}^-\right]\rho + \frac{\Gamma_1}{2}\mathcal{D}\left[\hat{Z}\otimes\hat{\sigma}^-\right]\rho. \label{eqn:BMRWA_ME2}
\end{align}
The first term on the right-hand-side of this master equation is exactly the uncorrelated decay of the spectator qubit that we want. However, the second term, which has an equal magnitude decay rate, is a multiplicative-correlated dephasing-decay process. 

This truly incoherent process is distinct from the spectator-decay-induced dephasing that is the focus of this manuscript. As we have discussed, the latter is a coherent but random phase kick on the control qubit that occurs when the spectator relaxes, which only appears incoherent in the ensemble average. The intuition behind the second term in Eq.~\eqref{eqn:BMRWA_ME2} is best understood from Eq.~\eqref{eqn:BMRWA_ME}, which consists of two decay channels that perfectly measure the state of the control qubit, resulting in the correlated dephasing of Eq.~\eqref{eqn:BMRWA_ME2}.

We see no evidence in our data that the system evolves according to Eq.~\eqref{eqn:BMRWA_ME2}, and so we are left to conclude that the BMRWA approach does not derive the correct model for the system. As we will see in the next section, it is the strong secular approximation that is to blame. The strong secular approximation captures the idea that the dissipation sets the slowest timescale of the system evolution, which is not necessarily valid for small $ZZ$-interaction where $1/\nu \sim T_1$. Thus, the BMRWA approach requires assumptions that do not hold for our system.

\subsection{Partial secular approximation: a successful derivation}

The BMPSA approach follows the same procedure as the BMRWA approach up until the Redfield equation, where it makes a partial secular approximation instead of the strong secular approximation of the RWA. The specific nature of this partial secular approximation, and the resulting master equation obtained, can take several forms, see e.g.~Refs.~\cite{Trushechkin21,McCauley20} and the references therein. However, the differences in form between different PSAs are small, e.g.~at what frequency is the bath spectral density evaluated to calculate the decay rate. Here, we will follow the approach of Ref.~\cite{Trushechkin21}.

The idea behind the partial secular approximation is to treat nearly degenerate Bohr frequencies as degenerate, and separate the Bohr frequencies into clusters $\{\Omega_i,\Omega_j,...\}$ that are contained within $\bar{\Omega} \pm \delta\Omega$, where $\bar{\Omega}$ is the average Bohr frequency and $\delta\Omega$ the spread of the cluster. Each cluster has a single Lindblad operator associated with it
\begin{align}
    \hat{L}_{\bar{\Omega}} = \sum_{\bar{\Omega} - \delta\Omega < \lambda_j - \lambda_k < \bar{\Omega} + \delta\Omega}\bra{\lambda_k}\hat{X}\ket{\lambda_j}\ketbra{\lambda_k}{\lambda_j},
\end{align}
and similar to the BMRWA, the master equation takes the form
\begin{align}
    \dot\rho = \sum_{\bar{\Omega}}\Gamma[\bar{\Omega}]\mathcal{D}[\hat{L}_{\bar{\Omega}}]\rho, \label{eqn:PSAME}
\end{align}
with the sum of dissipators now running over Bohr frequency clusters instead of individual Bohr frequencies.

For our two-qubit system, as $\nu$ is the small energy scale we have four Bohr frequency clusters $\{\omega_0+2\nu,\omega_0-2\nu\}$, $\{\omega_1+2\nu,\omega_1-2\nu\}$, $\{-\omega_0+2\nu,-\omega_0-2\nu\}$, and $\{-\omega_1+2\nu,-\omega_1-2\nu\}$. The resulting BMPSA master equation is
\begin{align}
    \nonumber\dot{\rho}  &= \Gamma(\omega_1)\mathcal{D}\left[\ketbra{1}\otimes\hat{\sigma}^- + \ketbra{0}\otimes\hat{\sigma}^-\right]\rho \\ &= \Gamma(\omega_1)\mathcal{D}\left[\hat{\mathbb{I}}\otimes\hat{\sigma}^-\right]\rho, \label{eqn:PSAME1}
\end{align}
where as before we have dropped terms proportional to $\abs{a}^2$ or with negative average Bohr frequencies, and $\Gamma(\omega_1)$ is defined as in Eq.~\eqref{eqn:decayrate}. For the two-qubit example, this matches Eq.~\eqref{eqn:targetME} for $\gamma_1 \equiv \Gamma(\omega_1)$.

Extending to multiple spectator qubits with independent baths, as in Eq.~\eqref{eqn:logHSB}, the master equation for each bath can be derived independently, following the procedure outlined above. While the physical qubit hybridization implies that a single bath may interact with multiple qubits, the interaction between each qubit and a given bath belongs to a distinct Bohr frequency cluster provided the qubits are all sufficiently detuned, such that $\abs{\omega_i - \omega_j} > \abs{\nu_{\rm max}}~\forall~i,j$, where $\nu_{\rm max}$ is the largest collective $ZZ$-shift due to the interaction with all other qubits. Thus, for the bath coupled to physical qubit $j$, by the PSA there will be no terms of order $\mathcal{O}(\abs{a_{jk}})$ in the master equation, and terms involving qubits $k\neq j$ will be order $\mathcal{O}(\abs{a_{jk}}^2)$.

Following these arguments, we arrive at a total master equation that has one term of the form of Eq.~\eqref{eqn:PSAME1} for each qubit
\begin{align}
    \nonumber\dot{\rho}  = \sum_{j=0}^N\Gamma_j(\omega_j)\mathcal{D}\left[\hat{\sigma}_j^-\right]\rho, \label{eqn:PSAME2}
\end{align}
which for $\gamma_j \equiv \Gamma_j(\omega_j)$ reproduces the relevant part of Eq.~\eqref{eqn:targetME}. This establishes both the justification, and appropriate assumptions required, for its use as the starting point of our theoretical modelling.

\subsection{Cumulant expansion and time coarse-graining}

While the BMPSA establishes the justification for our theoretical model, it does not offer a way of exploring the regime between its validity and the validity of the BMRWA. This becomes particularly relevant for multi-qubit systems. A Bohr frequency cluster centred around frequency $\omega_j$, corresponding to the decay of qubit $j$, will contain elements that differ in frequency by as much as four times the sum of the $ZZ$-interactions between qubits. Thus, the energy separation may approach the energy scale of the dissipation, calling into question the validity of the PSA as opposed to the RWA.

To bridge these two regimes, we consider a master equation derivation using the cumulant expansion and time coarse-graining (CETCG) approach \cite{Lidar19}. We follow exactly the approach described in Ref.~\cite{Lidar19}, and so we only outline the relevant points here. We consider the exact system-evolution propagator, which we describe as a the exponential of a power series in the system-bath coupling Hamiltonian known as its cumulant expansion. By a weak-coupling argument, we truncate this expansion to second order, and the first order term can be shown to vanish.

We then define a timescale, $\tau_C$, longer than the bath correlation timescale, $\tau_B$, but shorter than the relevant system timescales, $\tau_S$, i.e. $\tau_B\ll\tau_C\ll\tau_S$, and use this timescale to approximate the derivative of the system density matrix
\begin{align}
    \dot\rho \approx \frac{\rho(\tau_C) - \rho(0)}{\tau_C} = \mathcal{L}_{\tau_C}\rho.
\end{align}
Working this out explicitly with the cumulant expansion we can define the differential map $\mathcal{L}_{\tau_C}$, which can be shown to be of Lindblad-form \cite{Lidar19}. As in the other approaches we consider the CETCG derivation results in a time-independent $\mathcal{L}_{\tau_C}$, but introduces a ``user-defined'' timescale $\tau_C$ into the form of $\mathcal{L}_{\tau_C}$. In the limit $\tau_C \rightarrow \infty$ the CETCG master equation is equivalent to the BMRWA master equation.

Due to this additional timescale, the CETCG master equation has a more general structure than the BMRWA or BMPSA equations, given by
\begin{align}
    \dot\rho = \sum_{\Omega,\Omega'}\Gamma(\Omega,\Omega')\left(\hat{L}_\Omega\rho\hat{L}_{\Omega'}^\dagger - \frac{1}{2}\left\{\hat{L}_{\Omega'}^\dagger\hat{L}_\Omega,\rho\right\}\right), \label{eqn:CEME}
\end{align}
where, as before, $\Omega$ are the Bohr frequencies, $\hat{L}_\Omega$ are defined as in Eq.~\eqref{eqn:LinOps}, and
\begin{align}
    \nonumber&\Gamma(\Omega,\Omega') = \\ &\frac{1}{\tau_C}\int_0^{\tau_C} ds \int_0^{\tau_C} ds' e^{i(\Omega's - \Omega s')}\left<\hat{B}(s-s')\hat{B}(0)\right>. \label{eqn:CErate1}
\end{align}
Crucially, Eq.~\eqref{eqn:CEME} is not a diagonal Lindblad equation, and allows for coherent interaction between Lindblad operators of different Bohr frequencies. If $\Gamma(\Omega,\Omega') = 0~\forall~ \Omega\neq\Omega'$, then we obtain a master equation with the structure of the BMRWA derivation. If $\Gamma(\Omega,\Omega') = \Gamma(\Omega,\Omega) = \Gamma(\Omega',\Omega')$ within a Bohr frequency cluster, then we recover the structure of the BMPSA.

To more easily evaluate Eq.~\eqref{eqn:CErate1}, we rewrite it in terms of the coordinates $u=s-s'$ and $v=s+s'$ as
\begin{align}
    \nonumber&\Gamma(\Omega,\Omega') = \\ \nonumber&\frac{1}{\tau_C}e^{i\frac{\Omega'-\Omega}{2}\tau_C}\int_0^{\tau_C}dv\cos\left(\frac{\Omega'-\Omega}{2}(v-\tau_C)\right)\\
    &\times \int_{-v}^vdu~e^{i\frac{\Omega'+\Omega}{2}u}\left<\hat{B}(u)\hat{B}(0)\right>.
\end{align}
We first consider the case of $\Omega=\Omega'$, which after integration by parts reduces to the expression
\begin{align}
    \Gamma(\Omega,\Omega) = \int_{-\tau_C}^{\tau_C}du~e^{i\Omega u}\left<\hat{B}(u)\hat{B}(0)\right> + \mathcal{O}\left(\frac{\tau_B}{\tau_C}\right), \label{eqn:equal}
\end{align}
where we have only kept the leading order term (see Ref.~\cite{Lidar19} for the full expression). The more general case $\Omega\neq\Omega'$ can be expressed as
\begin{align}
    \nonumber&\Gamma(\Omega,\Omega') = \frac{e^{i\frac{\Omega'-\Omega}{2}\tau_C}}{(\Omega'-\Omega)\tau_C}\int_{-\tau_C}^{\tau_C}du\left<\hat{B}(u)\hat{B}(0)\right> \\ \nonumber&\times \Bigg(\sin\left(\frac{\Omega'-\Omega}{2}\tau_C\right)\left(e^{i\Omega u} + e^{i\Omega' u}\right) \\&-i~{\rm sgn}(u) \cos\left(\frac{\Omega'-\Omega}{2}\tau_C\right)\left(e^{i\Omega u} - e^{i\Omega' u}\right)\Bigg). \label{eqn:unequal}
\end{align}

As we have assumed that $\tau_B \gg \tau_C$, we can safely assume that the bath correlation function decays over a much shorter time scale than $\tau_C$, such that we can set the upper limits of the integrals in the above equations to $\infty$ (note that from here on we diverge from the analysis conducted in Ref.~\cite{Lidar19}). This results in
\begin{align}
    &\Gamma(\Omega,\Omega) \approx \Gamma(\Omega), \\
    &\nonumber\Gamma(\Omega,\Omega') = \frac{e^{i\frac{\Omega'-\Omega}{2}\tau_C}}{(\Omega'-\Omega)\tau_C}\Bigg(\sin\left(\frac{\Omega'-\Omega}{2}\tau_C\right)\left[\Gamma(\Omega) + \Gamma(\Omega')\right] \\ 
    &+ 2\cos\left(\frac{\Omega'-\Omega}{2}\tau_C\right)\left[S(\Omega) - S(\Omega')\right]\Bigg),
\end{align}
where $\Gamma(\Omega)$ and $S(\Omega)$ are the decay rate (defined as in Eq.~\eqref{eqn:decayrate}) and the Lamb shift (see Refs.~\cite{Breuer07,Lidar19} for definition) obtained by the BMRWA derivation. 

Notice that for $(\Omega'-\Omega)\tau_C \gg 1$ the overall factor in the denominator is such that $\Gamma(\Omega,\Omega') \approx 0$. For the two-qubit system we consider, this would be the case for Bohr frequencies that do not belong to the same cluster. In the opposite limit, assuming that the decay rate and Lamb shift do not vary significantly when $\Omega'-\Omega$ is small, as is the case within a Bohr frequency cluster, we can further simplify the rate expressions to
\begin{align}
    &\Gamma(\Omega,\Omega) \approx \Gamma(\Omega) \approx \Gamma(\bar{\Omega}), \label{eqn:GammaDiag} \\
    &\Gamma(\Omega,\Omega') \approx \Gamma(\bar{\Omega})e^{i\frac{\Omega'-\Omega}{2}\tau_C}{\rm sinc}\left(\frac{\Omega'-\Omega}{2}\tau_C\right). \label{eqn:GammaOffDiag}
\end{align}
We note that even though we have made an approximation by setting the limits of the integrals to infinity, the form of the above rates preserves positivity of the dynamics.

Plugging the above expressions into Eq.~\eqref{eqn:CEME} for our two-qubit system results in the master equation
\begin{align}
    \nonumber\dot{\rho}  &= \frac{\Gamma(\omega_1)}{2}\Bigg(\left(1 + {\rm sinc}\left(4\nu \tau_C\right)\right)\mathcal{D}\left[\hat{\mathbb{I}}\otimes\hat{\sigma}^-\right]\rho \\
    &+ \left(1 - {\rm sinc}\left(4\nu \tau_C\right)\right)\mathcal{D}\left[\hat{Z}\otimes\hat{\sigma}^-\right]\rho \label{eqn:CEME1} \\ 
    &\nonumber+ \sin(2\nu\tau_C){\rm sinc}\left(2\nu\tau_C\right)\left(i\hat{\mathbb{I}}\otimes\hat{\sigma}^-\rho\hat{Z}\otimes\hat{\sigma}^+ + h.c.\right)\Bigg). 
\end{align}
In the limit $\nu\tau_C \rightarrow 0$ this reduces to the BMPSA master equation of Eq.~\eqref{eqn:PSAME1}, and as previously mentioned, in the limit $\nu\tau_C \rightarrow \infty$ this becomes the BMRWA master equation of Eq.~\eqref{eqn:BMRWA_ME2}. Thus, by controlling the size of $\nu \tau_C$ we can interpolate between the two regimes of validity of our previous derivations, but by the properties of the CETCG derivation are guaranteed a physical master equation.

Unfortunately, as $\tau_C$ is ``user-defined'' and has no connection to physical observables we cannot use Eq.~\eqref{eqn:CEME1} to make quantitative predictions of experiments. However, we can make the qualitative prediction that as $\nu$ increases we expect to see the onset of a weak multiplicative-correlated dephasing-decay process, second line of Eq.~\eqref{eqn:CEME1}, and a corresponding reduction in the uncorrelated decay, first line of Eq.~\eqref{eqn:CEME1}. We could not have made such a qualitative prediction from either the BMRWA master equation, for which both rates are equal, or the BMPSA equation, for which there is no correlated dissipation.

Extending to multiple qubits, from the CETCG derivation we learn that we should expect weak multi-body correlated dissipation of the particular form that corresponds to dephasing on multiple qubits and decay on one, with Linblad operator $\bigotimes_{k \in \mathcal{S}}\hat{Z}_k \otimes \hat{\sigma}_j^-$ for $\mathcal{S}$ some subset of qubits. While we cannot quantitatively predict the rate of this dissipative process we know that it should scale with the spectator-qubit induced energy shift on the qubit that decays. 

In effect, we have determined the form of a new master equation to which we can fit experimental data, which is useful for two mains reasons. Firstly, identifying the likely form of dissipation is useful for designing error correction or error mitigation schemes, and correlated dissipation is particular important in this regard. Secondly, microscopic derivation can be used as a guide for ansatz building in tomography and in the design of protocols meant to measure specific dissipative rates. While we see no evidence for correlated dissipation in our current experiments, and thus no need to use the full Eq.~\eqref{eqn:CEME1} for modelling, the CETCG derivation gives us an immediate starting point were we to observe effects that could not be modelled by Eq.~\eqref{eqn:PSAME1} alone.

Finally, we note that in deriving Eq.~\eqref{eqn:CEME1} we have made the crucial assumption that $\Gamma(\Omega) = \Gamma(\Omega') = \Gamma(\bar\Omega)$ within a Bohr frequency cluster. If this is not the case, i.e.~the bath spectral density changes appreciably over the frequency scale defined by $\nu$, then this will also introduce a spatially correlated dephasing-decay process of the same form as the second line of Eq.~\eqref{eqn:CEME1}. Due to the fact that logical qubits are delocalized in space, temporal correlations in the noise (variations in bath spectral density) result in spatially correlated decoherence in the qubit array, even when the bath interacts locally with only one physical qubit.

\bibliography{SDID_bib}

\end{document}